\documentclass[smallextended,referee,envcountsect,]{svjour3}
\smartqed
\usepackage{graphicx}
\usepackage{chemformula}
\usepackage{longtable}
\usepackage{amssymb}
\usepackage{multirow,bigdelim}
\usepackage{caption}
\usepackage{subcaption}
\usepackage{cite}
\usepackage{enumitem}

\begin{document}

\title{Planning For Edge Failure in Fixed-Charge Flow Networks}


\author{Daniel Olson, Caleb Eardley, Sean Yaw}

\institute{Daniel Olson, Caleb Eardley, Sean Yaw (corresponding author) \at
             School of Computing \\
             Montana State University\\
              Bozeman, MT, USA\\
              sean.yaw@montana.edu
}


\maketitle

\begin{abstract}
The Fixed-Charge Network Flow problem is a well-studied NP-hard problem that has the goal of finding a flow in a network where fixed edge costs are incurred, regardless of the amount of flow hosted by the edge. 
In this paper, we consider scenarios where a designated edge in the network has the potential to fail after edges have already been purchased. 
If the edge does fail, procurement of additional edges may be required to repair the flow and compensate for the failed edge so as to maintain the original flow amount. 
We formulate a multi-objective optimization problem that aims to minimize the costs of both the initial flow as well as the repaired flow. 
We introduce an algorithm that finds the Pareto front between these two objectives, thereby providing decision makers with a sequence of solutions that trade off initial flow cost with repaired flow cost. 
We demonstrate the algorithm’s efficacy with an evaluation using real-world \ch{CO2} capture and storage infrastructure data.
\end{abstract}


\section{Introduction}

The basic Fixed-Charge Network Flow (FCNF) problem aims to find flows on a network where edges have a fixed cost incurred for all non-zero flow amounts~\cite{hirsch1968fixed}.
There are many variations of the basic FCNF problem formulation including each edge having a variable utilization cost (cost per unit of flow transported)~\cite{crainic2021fixed}, and capacitated~\cite{NIE202011231} or uncapacitated edges~\cite{BOURAS2019226}. 
The FCNF problem has been extensively studied in the context of many applications including transportation~\cite{akbari2020meta}, waste management~\cite{Kara2004}, civil infrastructure~\cite{willet2020water}, emergency response~\cite{biji2022model}, phased project planning~\cite{jones2022designing}, and railway management~\cite{laaziz2019service}. 

This paper studies the FCNF variant where edges have fixed costs, variable costs, and capacities, and the objective is to find the minimum cost flow such that the amount of flow equals a target value.
This paper considers how a flow should be determined if a designated edge in the network has the risk of failing.
If a flow is in operation on a network and then an edge fails, other edges may need to be opened to compensate for the closed edge in order to maintain the flow amount.
Since it is unknown if the designated edge will ultimately fail, two flow solutions need to be determined:
(1) An initial flow, in case the designated edge does not fail, and (2) a repaired flow in case the designated edge does fail.
The costs of these flows are dependent on each other, since the fixed costs of the failed edge, and any other edges abandoned by the flow when transitioning to the repaired flow, contribute to the cost of the repaired flow.
As such, determining a good initial flow and repaired flow is a multi-objective optimization problem with conflicting objectives. 
The Pareto front between these two objectives represents the solutions that decision makers should consider for trade-offs between decreasing initial cost or decreasing repaired cost.
The solution chosen from the Pareto front depends on the likelihood that the designated edge will fail and the decision maker's willingness to accept risk.
The novel contributions of this paper are threefold:
\begin{enumerate}
    \item The formulation of the multi-objective optimization problem aimed at minimizing the cost of the initial and repaired flows in an FCNF network.
    \item The introduction of a mixed-integer linear programming-based algorithm that finds the Pareto front between the cost of the initial flow and the cost of the repaired flow.
    \item An evaluation of this algorithm using real-world \ch{CO2} capture and storage infrastructure data to demonstrate its merit for decision makers.
\end{enumerate}

Though this paper is the first to explore finding the Pareto front between the initial flow cost and the repaired flow cost in an FCNF network, it is not the first research exploring edge failures in FCNF networks.
Many previous studies exist that aim to find \textit{survivable} networks, which are networks that are over-provisioned with edges so that a feasible flow will exist in the event of some failure scenario(s) occurring~\cite{Boginski2009, Anisi2018, Garg2008, Chen2013, eshghali2023risk, vspoljarec2021solving, Anisi2015, Anisi2019}.
Network survivability is a different objective than this paper, which aims to identify an initial and repaired solution, such that the repaired solution is only deployed in the event of edge failures and not preemptively.
Initially over-provisioning a network has strong resiliency benefits, but also incurs a cost that decision makers may not find appealing.

Other work considers multiple edge failures and aims to design flows that will limit the amount of value (as a function of flow) lost when edges fail~\cite{Sorokin2011}. 
The authors in the this work do not design flows with the aim for repairing them should a failure occur.
Similarly, in \cite{lotfi2021robust} the authors consider networks that can experience changes in edge capacity, amongst numerous other resiliency considerations. 
However, their objectives is to provision a network to handle demand fluctuations instead of repairing a network after edge failures.

The remainder of this paper is organized as follows:
Section~\ref{sec:back} introduces the problem's motivation. 
Section~\ref{sec:alg} formalizes the problem and presents a novel algorithm for finding the Pareto front. 
Section~\ref{sec:eval} presents an evaluation of the algorithm on real-world \ch{CO2} capture and storage infrastructure data. 
Section~\ref{sec:con} summarizes the contributions, discusses limitations, and proposes future research directions.

\section{Motivation}
\label{sec:back}
Finding a minimum cost flow in a flow network with fixed edge costs is a well-studied problem with many subtly different definitions.
For our version of the FCNF problem, we consider input consisting of a directed-edge graph with a designated source vertex, designated sink vertex, and a target flow value.
Each directed edge in the graph has a capacity, fixed cost to open the edge, and variable cost (cost per unit of flow) to utilize the edge.
If the amount of flow on an edge is non-zero, the cost incurred by that edge is its fixed cost plus its variable cost multiplied by the amount of flow on the edge.
As expected, if the amount of flow on the edge is zero, the cost incurred is also zero.
The objective of the FCNF problem is to find a valid flow of minimum cost whose flow value equals the target flow value.
This problem can be formulated as a mixed-integer linear program (MILP), as shown below:

\medbreak
\noindent
Instance Input Parameters:\\
\begin{tabular}{l l}
  \hspace{.5cm}$V$ & Vertex set\\
  \hspace{.5cm}$E$ & Edge set\\
  \hspace{.5cm}$s \in V$ & Source vertex\\
  \hspace{.5cm}$t \in V$ & Sink vertex\\
  \hspace{.5cm}$c_{(u,v)}$ & Capacity of edge $(u,v)$\\
  \hspace{.5cm}$a_{(u,v)}$ & Fixed opening cost of edge $(u,v)$\\
  \hspace{.5cm}$b_{(u,v)}$ & Variable utilization cost of edge $(u,v)$\\
  \hspace{.5cm}$T$ & Target flow amount
\end{tabular}

\medbreak
\noindent
Decision Variables:\\
\begin{tabular}{l l}
  \hspace{.5cm}$y_{(u,v)} \in \{0,1\}$ & Use indicator for edge $(u,v)$\\
  \hspace{.5cm}$f_{(u,v)} \in \mathbb{R}^{\ge 0}$ & Amount of flow on edge $(u,v)$
\end{tabular}

\medbreak
\noindent
Objective Function:
\begin{equation}
    \label{eq:obj}
    \min \sum_{(u,v) \in E} 
    \big(a_{(u,v)} y_{(u,v)} + b_{(u,v)} f_{(u,v)}\big)
\end{equation}

\medbreak
\noindent
Subject to the following constraints:
\begin{align}
\label{eq:Cap} &f_{(u,v)}\le c_{(u,v)} y_{(u,v)},\forall (u,v) \in E\\
\label{eq:Con} &\sum_{v \in V} f_{(u,v)} = \sum_{v \in V} f_{(v,u)}, \forall u \in V \setminus \{s,t\}\\
\label{eq:Tar} &\sum_{v \in V} f_{(s,v)} = T
\end{align}

\noindent
Where constraint~\ref{eq:Cap} enforces the capacity of each edge and forces $y_{(u,v)}$ to be set to one if $f_{(u,v)}$ is non-zero.
Constraint~\ref{eq:Con} enforces conservation of flow at each internal vertex.
Constraint~\ref{eq:Tar} ensures that the total flow amount meets the target.

Though the MILP described above will find an optimal flow when edges perform as parameterized, finding optimal solutions becomes more complicated when edges do not perform as parameterized.
In this paper, we consider scenarios where a designated edge may fail and become unusable after a flow on the flow network has already been identified and had its fixed costs paid for.
If the designated edge fails when it is part of a flow solution, the flow needs to be \textit{repaired} by finding an alternate $s-t$ route to compensate for the loss of the failed edge.
The cost of the repaired flow is its cost from Equation~\ref{eq:obj} plus the fixed cost of abandoned edges (including the failed edge) from the initial flow, since the fixed opening cost for those edges was already incurred prior to repairing the flow. 
Since it is not known if the designated edge will actually fail, and the fixed cost of each edge used in the initial solution needs to be paid even if it fails, determining a good initial flow is more complicated.
The problem we formulate is a multi-objective optimization problem that aims to minimize the cost of the initial flow and the cost of the repaired flow.
Since these objectives are conflicting, there is likely not one single optimal solution.
In fact, there can be a series of solutions that progressively trade off low initial cost for low repaired cost.

\begin{figure}[h]
\begin{center}
\includegraphics[scale=0.31]{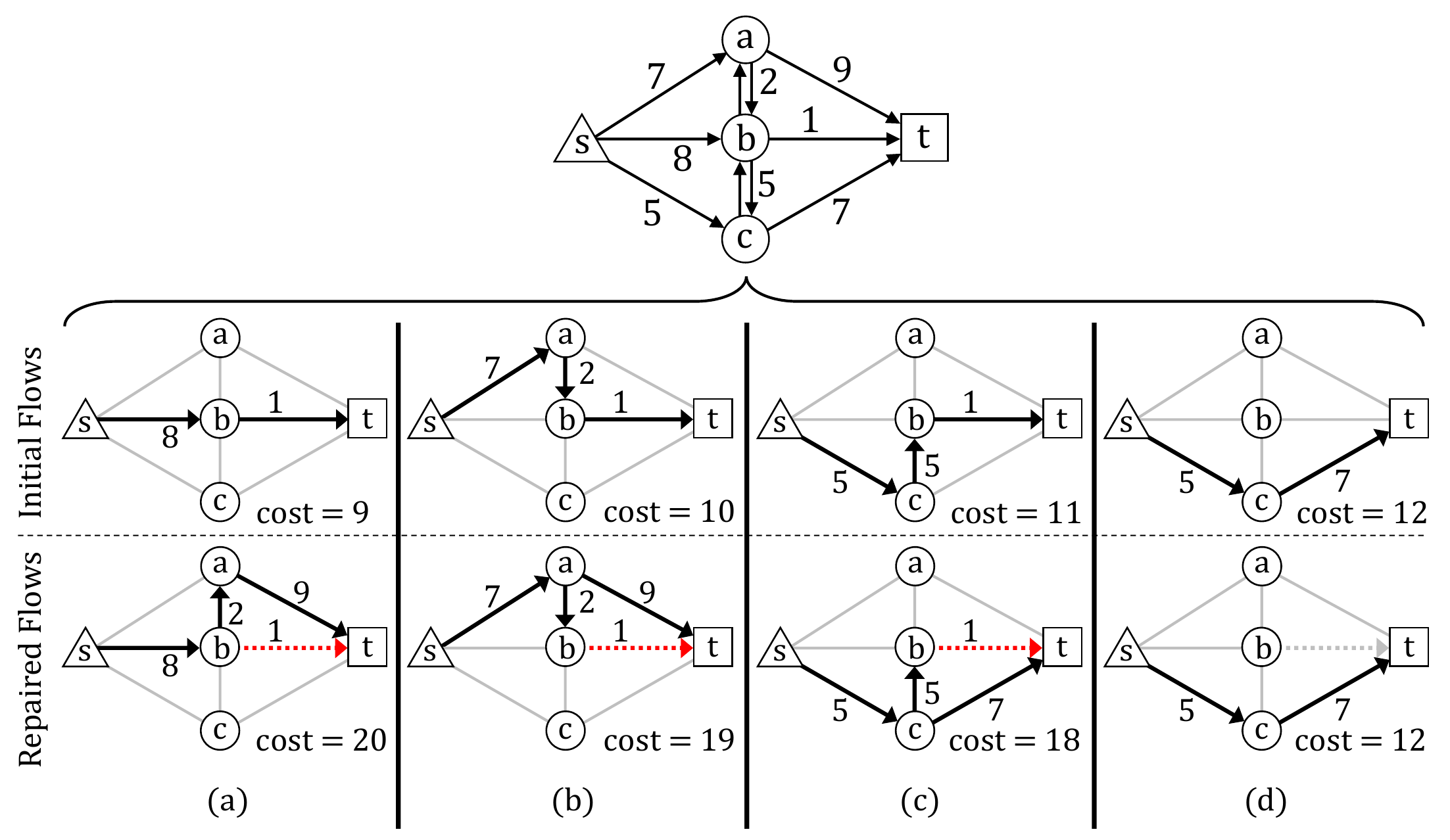}
\end{center}
\caption{Sample flow network (top) with source $s$, sink $t$, displayed fixed costs, no variable costs, unit edge capacities, and a target flow amount of one. Panels (a)-(d) show a sequence of initial flows (top) and optimal repaired flows (bottom) when edge $(b,t)$ fails.}
\label{fig:example}
\end{figure}

Consider the example flow network presented in Figure~\ref{fig:example} (top) where the edge weights shown are fixed costs, there are no variable costs, edge capacities all equal one, and the target flow amount is one.
In this scenario, the edge that could potentially fail is the edge $(b,t)$.
The minimum cost $s-t$ flow (Figure~\ref{fig:example}a top) is the path $(s,b,t)$ whose cost is $9$.
If the edge $(b,t)$ does not fail, this would be the uncontested optimal solution.
However, if this flow is deployed and $(b,t)$ fails, the cheapest way to repair this flow (Figure~\ref{fig:example}a bottom) requires a total cost of $20$.
In comparison, the three remaining solutions presented in panels (b)-(d) require progressively higher initial flow costs (top), but progressively lower repaired flow costs (bottom).
Depending on the confidence that the edge $(b,t)$ will not fail, the best initial flow (and corresponding repaired flow) can be determined from this set of possible solutions.

The sequence of solutions that would be of utility to a decision maker is the Pareto front between the initial cost and the repaired cost.
This would define a sequence of Pareto-optimal solutions where increasing initial flow cost corresponds to decreasing repaired flow cost.
The first solution in the sequence has an initial flow that is the minimum cost flow on the flow network (Figure~\ref{fig:example}a top), since this is the optimal choice if the designated edge does not fail.
Conversely, the final solution in the sequence has an initial flow and repaired flow that is the minimum cost flow on the flow network with the designated edge excluded (Figure~\ref{fig:example}d), since this is the optimal choice if the designated edge does fail.
The intermediate solutions are alternative options that trade off initial flow cost with repaired flow cost.
Pursuing a lower cost initial flow  with its associated higher cost repaired flow would make sense if the designated edge is unlikely to fail, as an expensive repaired flow has no bearing on the cost of the initial flow.
Pursuing a higher cost initial flow with its associated lower cost repaired flow corresponds to overbuilding the initial flow so that the repaired flow will be less expensive, should the designated edge fail.
If the edge is likely to fail, it could make sense to overbuild the initial flow enough to reduce the cost of repairing it, as it is likely that the repaired flow will be used.
The goal of this paper is to present an algorithmic process for finding the Pareto front between the initial flow cost and the repaired flow cost.

\section{Algorithm}
\label{sec:alg}
In this section, we present an algorithm that finds the Pareto front between the initial flow cost and the repaired flow cost.
This sequence of solutions has progressively decreasing repaired flow costs, starting with the minimum cost flow and proceeding to the minimum cost flow on the network with the designated edge excluded.
The problem input we consider is very similar to the input to the MILP described in Section~\ref{sec:back}, with the addition of the designated edge that may or may not fail:

\begin{tabular}{l l}
  \hspace{.5cm}$V$ & Vertex set\\
  \hspace{.5cm}$E$ & Edge set\\
  \hspace{.5cm}$s \in V$ & Source vertex\\
  \hspace{.5cm}$t \in V$ & Sink vertex\\
  \hspace{.5cm}$c_{(u,v)}$ & Capacity of edge $(u,v)$\\
  \hspace{.5cm}$a_{(u,v)}$ & Fixed cost of edge $(u,v)$\\
  \hspace{.5cm}$b_{(u,v)}$ & Variable cost of edge $(u,v)$\\
  \hspace{.5cm}$T$ & Target flow amount\\
  \hspace{.5cm}$(w,x) \in E$ & Edge subject to failure
\end{tabular}
\medbreak

The algorithm presented here works by solving a series of three MILPs that identify the sequence of solutions that form the Pareto front.
The first solution in the sequence is the minimum cost flow and the minimum cost repair of that flow when the designated edge fails. 
This solution corresponds to the initial and repaired flows shown in Figure~\ref{fig:example}a.
The three MILPs, run sequentially, each concurrently find an initial flow with the designated edge not failing and a repaired flow with the designated edge failing using the following decision variables:

\begin{tabular}{l l}
  \hspace{.5cm}$y^i_{(u,v)} \in \{0,1\}$ & Use indicator for edge $(u,v)$ in initial flow\\
  \hspace{.5cm}$f^i_{(u,v)} \in \mathbb{R}^{\ge 0}$ & Amount of flow on edge $(u,v)$ in initial flow\\
  \hspace{.5cm}$y^r_{(u,v)} \in \{0,1\}$ & Use indicator for edge $(u,v)$ in repaired flow\\
  \hspace{.5cm}$f^r_{(u,v)} \in \mathbb{R}^{\ge 0}$ & Amount of flow on edge $(u,v)$ in repaired flow
\end{tabular}
\medbreak

There is significant commonality in the constraints across the three MILPs.
These common constraints enforce the validity of the initial and repaired flows, and encode the failure of the designated edge for the repaired flow:

\begin{align}
  \ldelim\{{6}{*}[Initial Flow] \label{eq:iCap} &  f^i_{(u,v)}\le c_{(u,v)} y^i_{(u,v)},\forall (u,v) \in E &&\\
  \label{eq:iCon} &\sum_{v \in V} f^i_{(u,v)} = \sum_{v \in V} f^i_{(v,u)}, \forall u \in V \setminus \{s,t\}\\
  \label{eq:iTar} &\sum_{v \in V} f^i_{(s,v)} = T&&\\
  \nonumber\\
  \ldelim\{{6}{*}[Repaired Flow] \label{eq:rCap} &f^r_{(u,v)}\le c_{(u,v)} y^r_{(u,v)},\forall (u,v) \in E&&\\
  \label{eq:rCon} &\sum_{v \in V} f^r_{(u,v)} = \sum_{v \in V} f^r_{(v,u)}, \forall u \in V \setminus \{s,t\}\\
  \label{eq:rTar} &\sum_{v \in V} f^r_{(s,v)} = T&&\\
  \nonumber\\
  \ldelim\{{2.5}{*}[Edge Failure] \label{eq:fail} &f^r_{(w,x)} = 0&&\\
  \label{eq:fixed} &y^r_{(u,v)} \ge y^i_{(u,v)}&&
\end{align}
\medbreak

\noindent
Where Constraints~\ref{eq:iCap} and \ref{eq:rCap} restrict the flow amount on each edge to its capacity. 
Constraints~\ref{eq:iCon} and \ref{eq:rCon} enforce conservation of flow at each internal vertex.
Constraints~\ref{eq:iTar} and \ref{eq:rTar} ensure that the total flow amounts in the initial and repaired flows meet the target. 
Constraint~\ref{eq:fail} ensures that there is no flow on the failed edge in the repaired flow. 
Constraint~\ref{eq:fixed} ensures that if an edge is selected in the initial flow, it must also be selected in the repaired flow. 
This constraint is used to force the repaired flow to pay the fixed costs of any edges purchased for the initial flow but not used by the repaired flow, including the failed edge.

The algorithm functions in an iterative fashion, at each iteration finding an initial and repaired flow where the cost of the repaired flow is smaller than the cost of the repaired flow from the previous iteration. 
It is important to actively decrease the cost of the repaired flow instead of actively increasing the cost of the initial flow.
Finding an initial flow with the same cost as the preceding iteration but with a lower cost repaired flow is a useful solution for the decision maker to consider.
Conversely, an initial flow with an increased cost compared to the preceding iteration and the same repaired flow cost is of little utility.
At each iteration of the algorithm, three MILPs are used to identify the next initial and repaired flows in the sequence.
An overview of the algorithm is presented in Figure~\ref{fig:alg} and the three MILPS are detailed below:

\begin{figure}[h]
    \centering
    \includegraphics[scale=0.34]{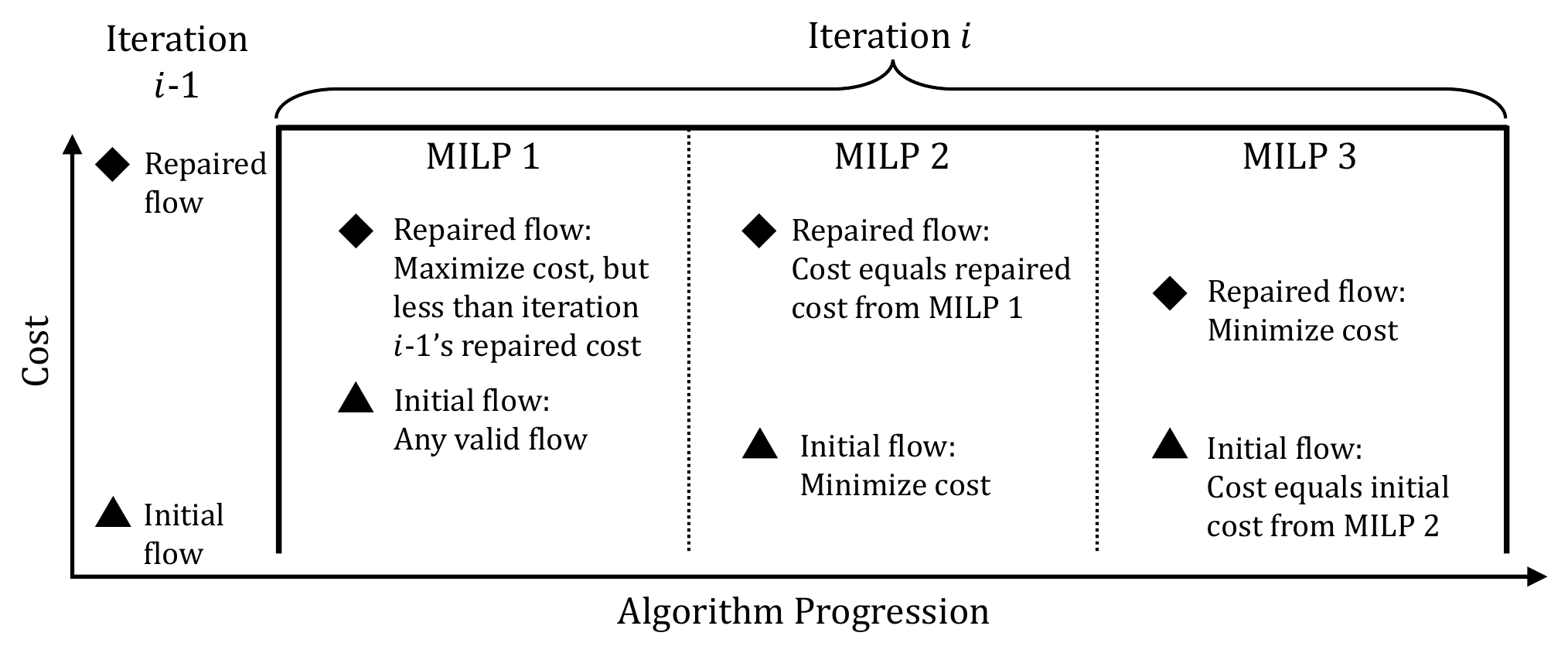}
    \caption{Overview of the objectives that the three MILPs optimize for when finding an initial and repaired flow during a single iteration of the algorithm. The initial and repaired flows found by MILP 3 form the next solution on the Pareto front.}
    \label{fig:alg}
\end{figure}

\begin{enumerate}[leftmargin=*, labelindent=0pt, label=MILP \arabic*.]
    \item First, an initial and repaired flow are found so that the cost of the repaired flow is maximized and the cost of the repaired flow is less than the cost of the previous iteration's repaired flow. 
    These flows can be found by formulating an MILP with the same input parameters and decision variables described earlier in this section and the following objective function that maximizes the cost of the repaired flow:
    \begin{equation}
    \nonumber
    \max \sum_{(u,v) \in E} 
    \big(a_{(u,v)} y^r_{(u,v)} + b_{(u,v)} f^r_{(u,v)}\big)
    \end{equation}
    The constraints for the MILP are the same as the constraints described earlier in this section with the addition of a constraint that requires the cost of the repaired flow to be less than the cost of the repaired flow from the last iteration. Let $R^{last}$ be the cost of the repaired flow from the last iteration and $\epsilon$ be some small value that represents the minimum step size for each sequential solution:
    \begin{equation}
    \nonumber
    \sum_{(u,v) \in E} 
    \big(a_{(u,v)} y^r_{(u,v)} + b_{(u,v)} f^r_{(u,v)}\big) \le R^{last} - \epsilon
    \end{equation}
    Maximizing the cost of a flow of a designated amount can result in undesired MILP behavior such as pushing flow along a cycle (thereby increasing the cost but not violating conservation of flow) or choosing intentionally expensive routes.
    This means that the repaired flow found is not necessarily the next one in the solution sequence, as its cost may be artificially inflated.
    Regardless, this MILP identifies the maximum cost at which there exists a valid repaired flow (and an accompanying initial flow) that is cheaper than the previous iteration's repaired flow.
    \medbreak

    \item Second, the cheapest initial flow is found such that its accompanying repaired flow equals the cost of the repaired flow found by MILP 1.
    As in the previous step, these flows are found by formulating an MILP with the common input parameters, decision variables, and constraints listed earlier in this section.
    The following objective function minimizes the cost of the initial flow:
    \begin{equation}
    \nonumber
    \min \sum_{(u,v) \in E} 
    \big(a_{(u,v)} y^i_{(u,v)} + b_{(u,v)} f^i_{(u,v)}\big)
    \end{equation}
    An additional constraint is needed that forces the cost of the repaired flow to equal the cost of the repaired flow found by MILP 1 ($R^{step1}$):
    \begin{equation}
    \nonumber
    \sum_{(u,v) \in E} 
    \big(a_{(u,v)} y^r_{(u,v)} + b_{(u,v)} f^r_{(u,v)}\big) = R^{step1}
    \end{equation}
    The result of this step is the cheapest initial flow whose corresponding repaired flow is cheaper than the previous iteration's repaired flow.
    All that is left to address is the issue that the cost of the repaired flow from step one may be artificially inflated. 
    \medbreak
    
    \item Third, an initial and repaired flow are found so that the cost of the repaired flow is minimized and the cost of the initial flow equals the cost of the initial flow found by MILP 2.
    These flows are found by an MILP with the common input parameters, decision variables, and constraints listed earlier in this section.
    The following objective function minimizes the cost of the repaired flow:
    \begin{equation}
    \nonumber
    \min \sum_{(u,v) \in E} 
    \big(a_{(u,v)} y^r_{(u,v)} + b_{(u,v)} f^r_{(u,v)}\big)
    \end{equation}
    An additional constraint is needed that forces the cost of the initial flow to equal the cost of the initial flow found by MILP 2 ($I^{step2}$):
    \begin{equation}
    \nonumber
    \sum_{(u,v) \in E} 
    \big(a_{(u,v)} y^i_{(u,v)} + b_{(u,v)} f^i_{(u,v)}\big) = I^{step2}
    \end{equation}
    The result of this step is the next initial and repaired flow in the solution sequence.
\end{enumerate}

This algorithm generates a sequence of initial and repaired flows that define the Pareto front.
The algorithm is initialized so that the initial flow for iteration $0$ is the minimum cost flow and the repaired flow is the minimum cost repair of the initial flow when the designated edge fails.
It is initialized to this because it is the cheapest initial flow, where no restrictions are put on the cost of the repaired flow.
The final iteration of the algorithm finds a solution where the initial flow and repaired flow are both the minimum cost flow on the flow network with the designated edge excluded.
It finds this solution because it is the cheapest possible repaired flow.
The intermediate solutions all represent trade-offs between initial flow cost and repaired flow cost.

\section{Evaluation}
\label{sec:eval}
\begin{figure}[h]
    \centering
    \includegraphics[scale=0.34]{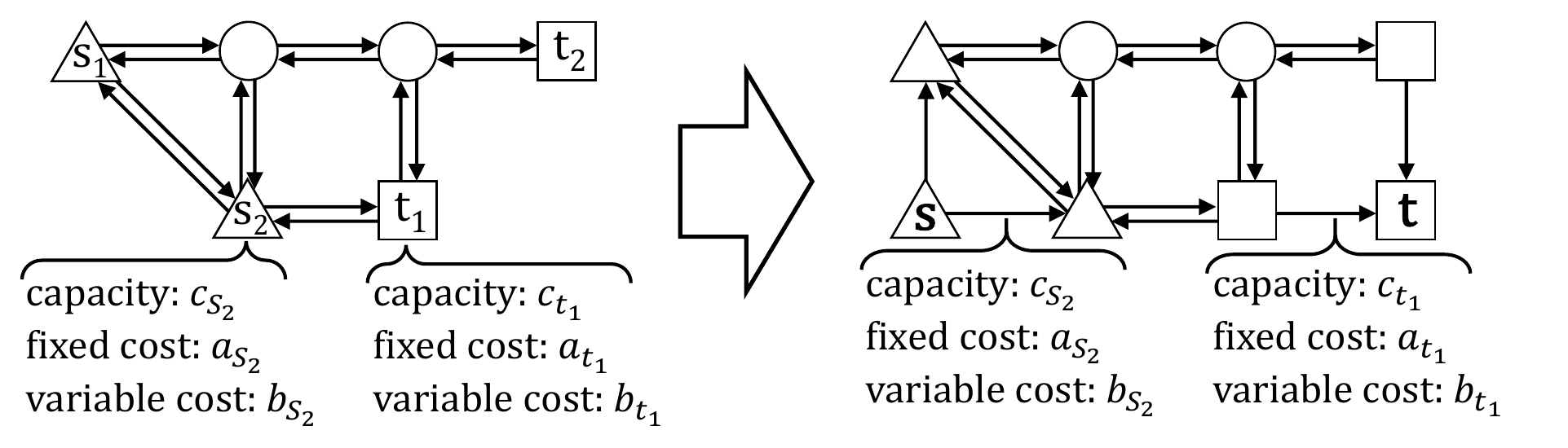}
    \caption{A sample reduction from a CID problem instance (left) to a FCNF instance (right).
    A new source vertex is added to the FCNF instance that has directed edges from it to each of the CID instance source vertices.
    A new sink vertex is added that has directed edges to it from each of the CID instance sink vertices.
    The capacities and costs of the source and sink vertices in the CID instance are assigned to the new directed edges in the FCNF instance.
    The capacities and costs of the edges in the original CID instance have the same values in the FCNF instance.
    Non-source and non-sink vertices in CID instances do not have capacities or costs.
    This demonstrates that the CID problem is identical to the FCNF problem.}
    \label{fig:reduct}
\end{figure}

In this section, we present results from running the algorithm described in Section~\ref{sec:alg} on \ch{CO2} capture and storage (CCS) infrastructure data.
CCS is a cornerstone climate change mitigation strategy that works by capturing \ch{CO2} from industrial emissions (sources), transporting it through a pipeline network (edges), and injecting it into geologic reservoirs (sinks) for long-term storage. 
The CCS Infrastructure Design (CID) problem aims to determine the most cost-effective sources, sinks, and edges to open in order to process a pre-determined quantity of \ch{CO2}~\cite{whitman2021scalable}.
Sources, sinks, and possible pipeline edges are all parameterized as having a capacity, fixed cost, and variable utilization cost.
For sources, the capacity is the annual amount of \ch{CO2} emitted by that source, the fixed cost is the cost to retrofit the source for capture operations, and the variable cost is the cost per ton of \ch{CO2} to actually capture the \ch{CO2}.
For sinks, the capacity is the annual amount of \ch{CO2} that can be injected into the reservoir, the fixed cost is the cost to construct the injection well, and the variable cost is the cost per ton of \ch{CO2} to inject the \ch{CO2}.
For possible pipeline edges, the capacity is the annual amount of \ch{CO2} that can be transported in that pipeline, the fixed cost is the cost to construct the pipeline, and the variable cost is the cost per ton of \ch{CO2} to transport the \ch{CO2}.
The CID problem is identical to the FCNF problem as shown in Figure~\ref{fig:reduct}.

Large uncertainties often surround the capacity and injection cost of sinks. 
Even with thorough site characterization, the true performance of a sink is not well known until injection operations commence. 
If a sink underperforms after infrastructure has been deployed, additional cost will need to be incurred to provision the project with more injection wells, pipeline capacity, or pipelines routes to better storage formations, or the success of the CCS project will be put at risk. 
The performance uncertainty associated with capture and transport is far less than storage in CCS projects.
As such, for the purpose of this evaluation we consider sink locations as the designated resources that may or may not fail in the FCNF network.

\begin{figure}[h]
     \centering
     \begin{subfigure}[t]{0.49\textwidth}
         \centering
         \includegraphics[scale=0.35]{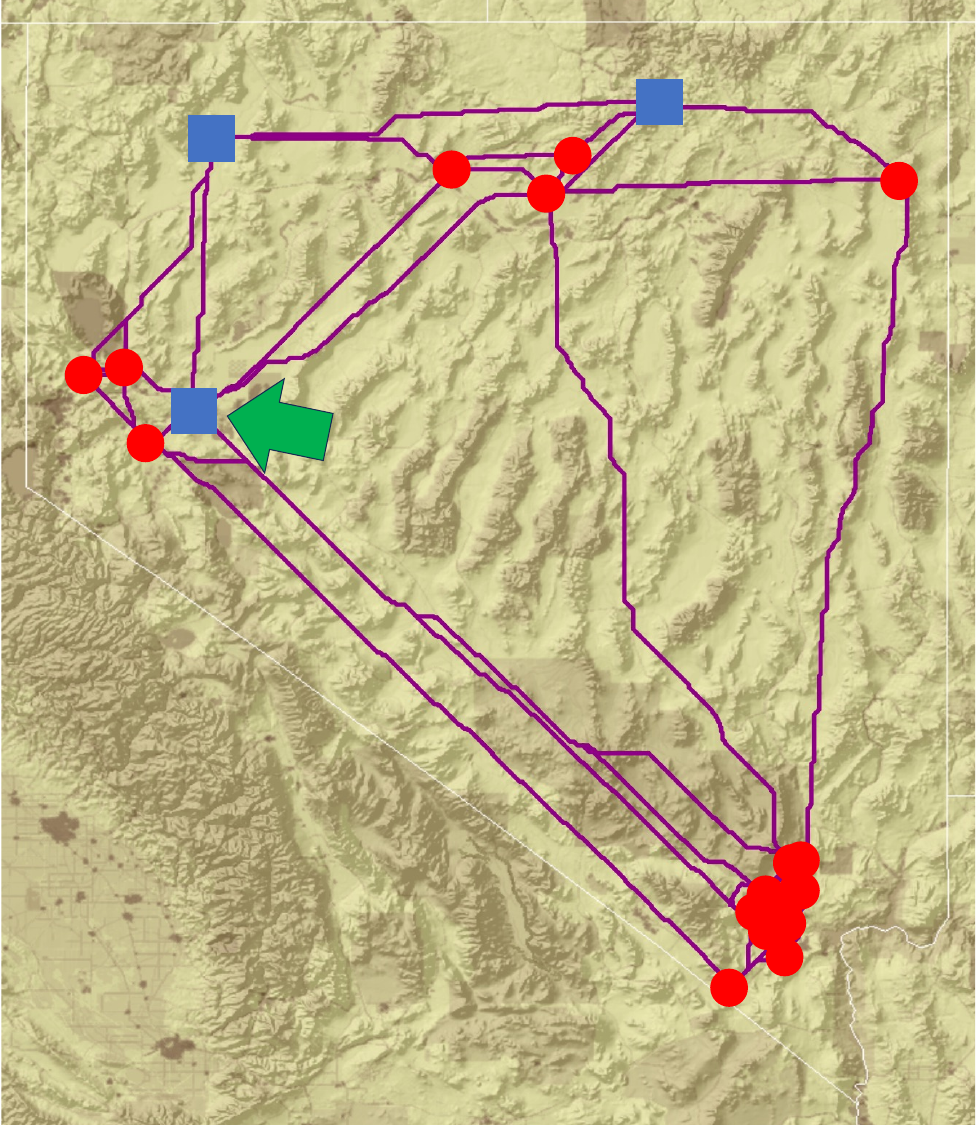}
         \caption{CCS data consisting of sources (circles), sinks (squares), and possible pipeline routes (edges). The arrow indicates the sink that fails in the evaluated scenario.}
         \label{fig:data}
     \end{subfigure}
     \hfill
     \begin{subfigure}[t]{0.49\textwidth}
         \centering
         \includegraphics[scale=0.35]{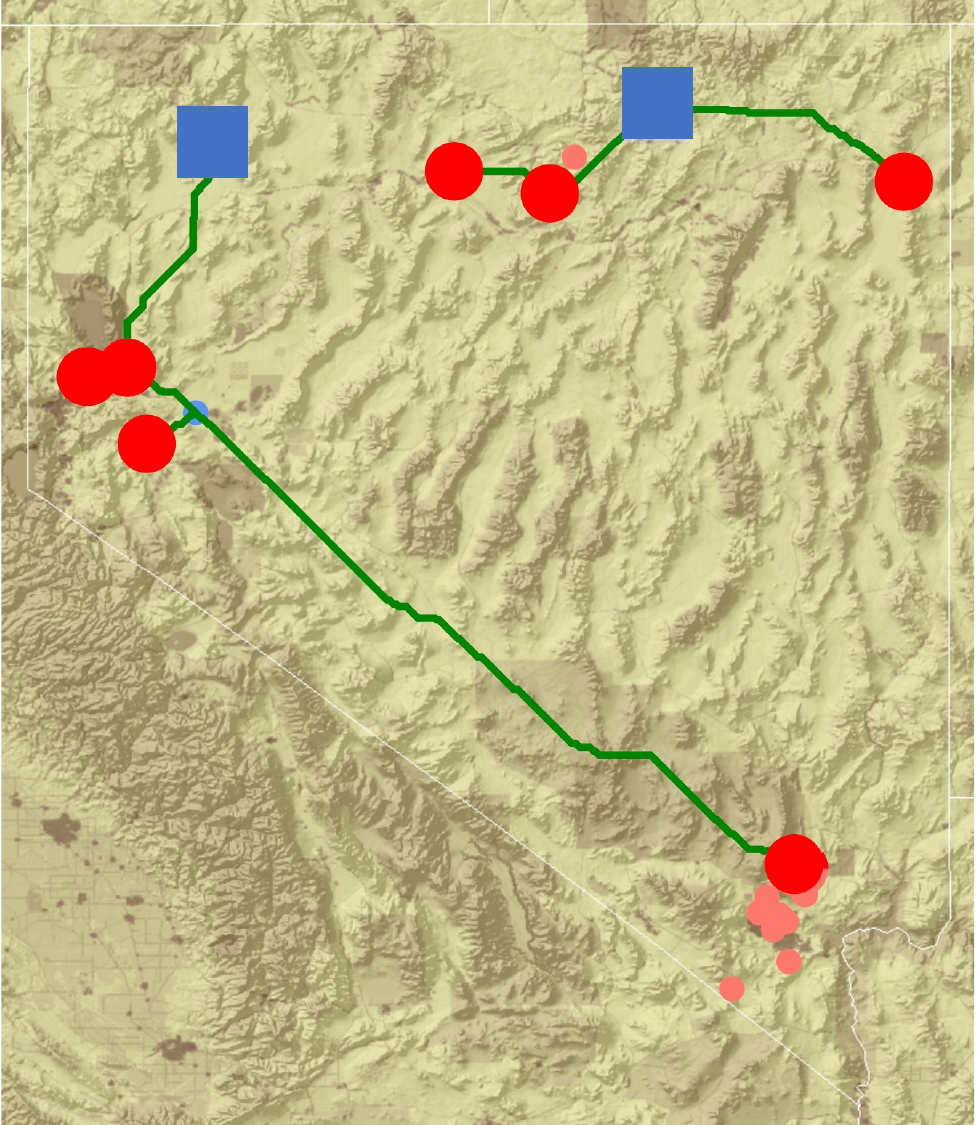}
         \caption{Example of a solved CCS instance with selected sources (large circles), sinks (large squares), and deployed pipelines (edges).}
         \label{fig:solved}
     \end{subfigure}
        \caption{Nevada CCS data set and sample solution.}
        \label{fig:maps}
\end{figure}

The data we use for this evaluation was collected as part of the US Department of Energy's (DOE) Carbon Utilization and Storage Partnership project, one of the DOE’s Regional Initiatives to Accelerate CCS Deployment.
We consider a data set consisting of 21 sources, three sinks, and 5230 km of possible pipeline routes within the US state of Nevada.
The total annual emission amount in this dataset is $15.92$ million tons of \ch{CO2} per year.
The total storage capacity in this dataset is $392.3$ million tons of \ch{CO2}.
Figure~\ref{fig:data} shows the location of the sources (red vertices), sinks (blue vertices), and possible pipeline routes (purple edges).
In this evaluation, we considered the southernmost sink (indicated with the arrow) as the sink that may fail.
The algorithm from Section~\ref{sec:alg} was implemented into the CCS infrastructure modelling tool $SimCCS$ which utilizes an MILP to concurrently determine cost-minimal capture, transport, and storage infrastructure~\cite{Middleton2020}.
$SimCCS$ also provides a standardized way to load, manipulate, and visualize CCS data.
We replaced the $SimCCS$ MILP core with the algorithm from Section~\ref{sec:alg}, but retained the data loading, manipulation, and visualization capabilities.
An example of a solved CID instance as viewed in $SimCCS$ is shown in figure~\ref{fig:solved}.

To make the results more applicable to CCS applications, an additional constraint was added to the MILPs in the model that requires the capture amount at each source in the repaired flow to equal what was captured at that source in the initial flow.
The reason for this is that this specific CCS dataset does not have a fixed cost to retrofit the emitter for capture operations.
Instead, the variable capture cost identified for each site assumes the site is in operation for the full project length, which will account for the necessary retrofit cost.
This results in the cost of modeled solutions being more realistic if the same capture facilities are utilized both before and after failure.

\begin{figure}[h]
    \centering
    \includegraphics[scale=0.5]{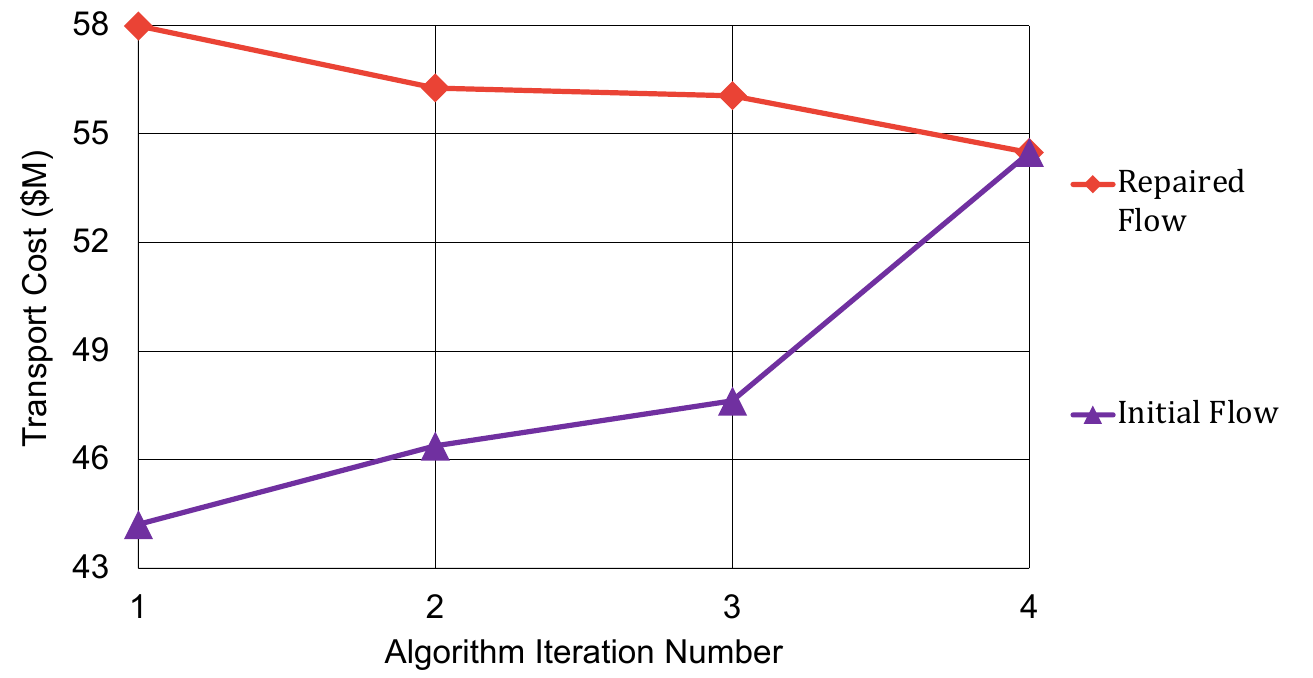}
    \caption{Progression of the initial and repaired flows,  over the course of the execution of the algorithm, from the minimum cost flow (and its associated repaired flow) to minimum cost flow on the network with sink number two excluded.}
    \label{fig:progression}
\end{figure}

A scenario was run consisting of capturing $8$ million tons of \ch{CO2} per year for $20$ years while identifying sink number two (arrow in Figure~\ref{fig:data}) as the sink that may fail.
Figure~\ref{fig:progression} shows the transport cost of the initial and repaired flows over the course of four iterations of the algorithm.
Only the transport cost is shown since the capture cost is the same between initial and repaired flows (due to the additional constraint) and the storage cost is the same because the dataset has identical storage costs for all storage locations.
As designed, the initial flow for the first iteration is the minimum cost flow on the network, and its repaired flow is the repaired minimum cost flow.
The initial and repaired flows of the final iteration are both the minimum cost flow on the flow network with sink number two excluded. 
The initial and repaired flows found in the intermediate iterations demonstrate a trade-off between increasing initial flow cost and decreasing repaired flow cost.

\begin{figure}[h]
    \centering
    \includegraphics[scale=0.5]{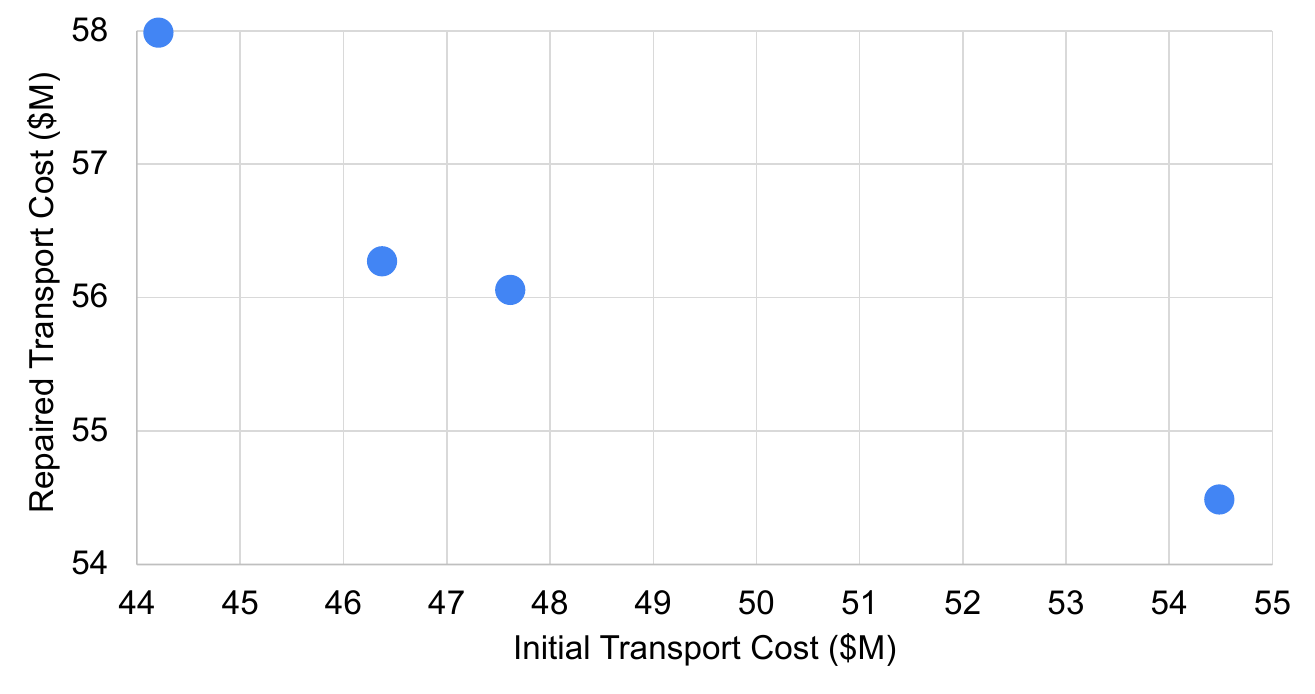}
    \caption{Pareto front between initial transport cost and repaired transport cost.}
    \label{fig:pareto}
\end{figure}

Figure~\ref{fig:pareto} shows the sequence of Pareto-optimal points, forming the Pareto front, between the initial flow transport cost and the repaired flow transport cost found by the algorithm for the scenario capturing $8$ million tons of \ch{CO2} per year for $20$ years. 
These solutions represent the options for optimal trade-offs between initial flow cost and repaired flow cost.
These are the solutions that a decision maker would take into account when determining the likelihood of the designated edge failing, the cost of such a failure, and their willingness to accept risk.
This evaluation demonstrates the ability of our algorithm to identify the Pareto front in a relevant application with real-world data. 

\section{Conclusions}
\label{sec:con}
This paper introduces a variation of the FCNF problem in which the possibility of failure for a specified edge in the network exists.
This suggests the objective of minimizing both the initial flow cost as well as the post-failure repaired flow cost.
An algorithm is presented that identifies the Pareto optimal points comprising the Pareto front between the initial flow cost and repaired flow cost.
These points enable decision makers to identify trade-offs between initial cost and repaired cost.
The algorithm functions by solving three MILPs at each iteration to identify one of the Pareto optimal points.
The algorithm was evaluated on real-world CCS data and demonstrated its efficacy at identifying the Pareto front.

There are multiple avenues of future work on this problem and algorithm. 
First, the algorithm currently only addresses the possible failure of a single pre-determined edge.
It is more likely that, in a practical application, there would be multiple edges at risk of failure, with different failure risks.
A process that identifies this higher-dimension Pareto front and provides tools for navigating it would likely be beneficial to decision makers. 

Secondly, given the presented algorithm's dependence on solving multiple MILPs, future work could include generating a process for finding a high-quality, sub-optimal Pareto front in polynomial time.
The evaluation presented in this paper demonstrates that the algorithm is useful for real-world problems, but MILP intractability limitations would prevent this algorithm from being used on large-scale problems with hundreds of edges. 
Developing algorithms that are not dependent on solving MILPs could be of significant value to certain application areas.

\begin{acknowledgements}
This research was funded by the U.S. Department of Energy's Fossil Energy Office through the Carbon Utilization and Storage Partnership (CUSP) for the Western USA (Award No. DE-FE0031837).
\end{acknowledgements}

\par\addvspace{17pt}\small\rmfamily
\trivlist\if!\ackname!\item[]\else
\item[\hskip\labelsep
{\bfseries Data Availability}]\fi
The data referenced in this study was generated by private and public partners as part of the U.S. Department of Energy's Carbon Utilization and Storage Partnership (CUSP). The corresponding author can submit requests to share the data with interested parties on reasonable request.

\bibliographystyle{spmpsci} 
\bibliography{refs}

\end{document}